\documentclass[a4paper,fleqn,usenatbib]{mnras}
\usepackage{newtxtext,newtxmath}

\usepackage[T1]{fontenc}
\usepackage{ae,aecompl}

\usepackage{graphicx}	
\usepackage{amsmath}	
\usepackage{amssymb}	

\title[Light curve variation]{Light curve variation caused by
 accretion column switching stellar hemispheres}

\author[M. \v{C}emelji\'{c}]{
Miljenko \v{C}emelji\'{c}$^{1}$\thanks{E-mail: miki@camk.edu.pl (M\v{C})}
\\
$^{1}$Nicolaus Copernicus Astronomical Center, Bartycka 18, 00-716
Warsaw, Poland\\
}

\date{Accepted XXX. Received YYY; in original form ZZZ}

\pubyear{2018}

\begin{document}
\label{firstpage}
\pagerange{\pageref{firstpage}--\pageref{lastpage}}
\maketitle

\begin{abstract}
I use two-dimensional axisymmetric numerical simulations of star-disk
magnetospheric interaction to construct a three-dimensional model of
hot spots on the star created by infalling accretion columns.
The intensity of emitted radiation is computed with minimal assumptions,
as seen by observers at infinity from the different angles. In the
illustrative examples, shown is a change in the intensity curve
with changes in the geometry of the model, and also a change in results
with modification in the physical parameters in the simulation.
\end{abstract}



\section{Introduction}
\begin{figure}
\includegraphics[width=\columnwidth]{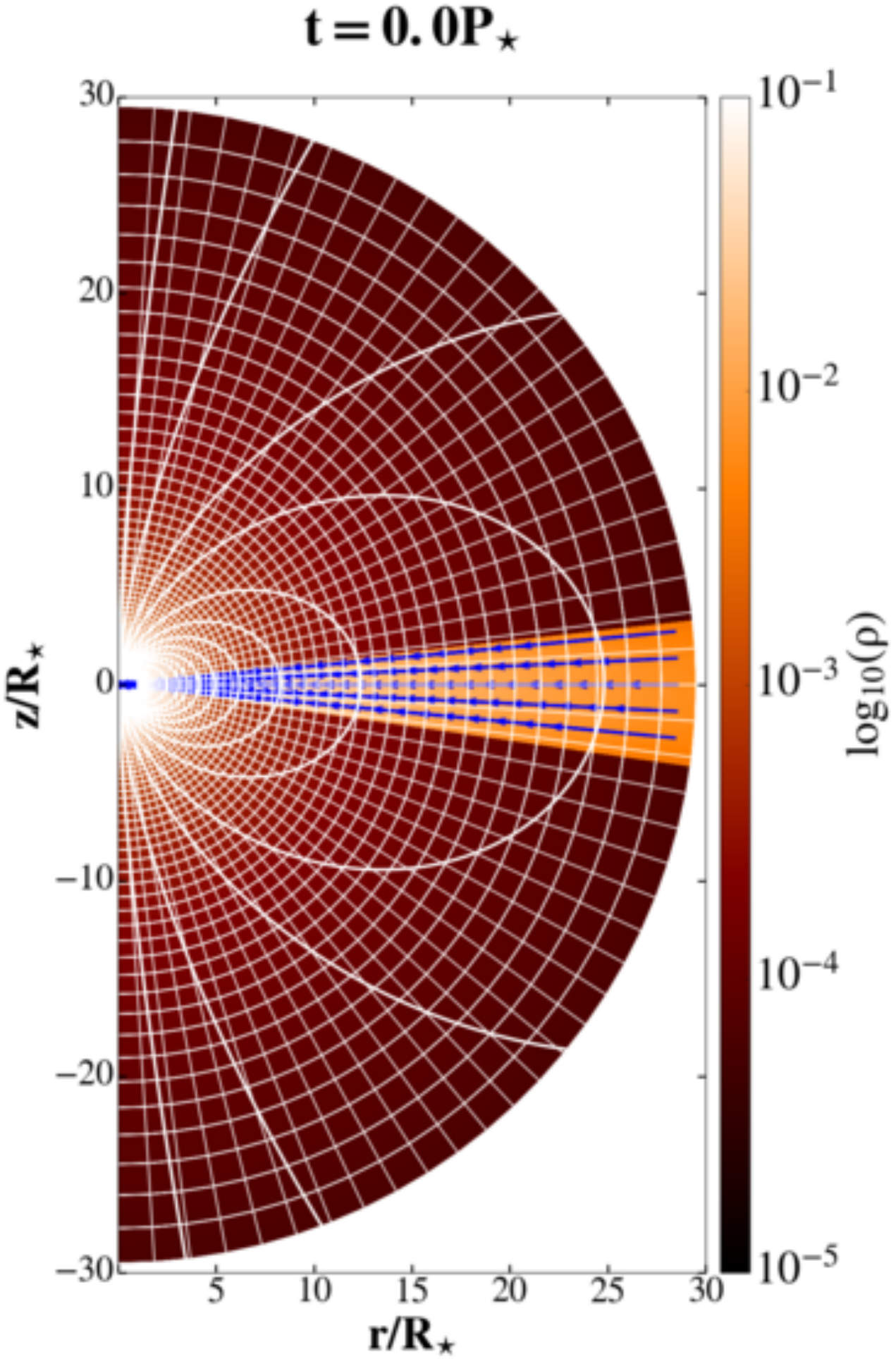}
\caption{Initial density in the accretion disk and stellar corona in a
hydrostatic equilibrium is shown in logarithmic color grading in code units,
in the case with $\theta=[0,\pi]$. A sample of initial velocity vectors and
stellar dipole magnetic field lines (solid lines) is shown, together with
the computational grid represented with a 4x4 block of cells. In the
$\theta=[0,\pi/2]$ case, initial setup is identical to the top quadrant
shown here, with the addition of the symetric boundary conditions at the
disk equatorial plane.}
\label{fulpi}
\end{figure}
\begin{figure*}
\centering
\includegraphics[width=0.67\columnwidth]{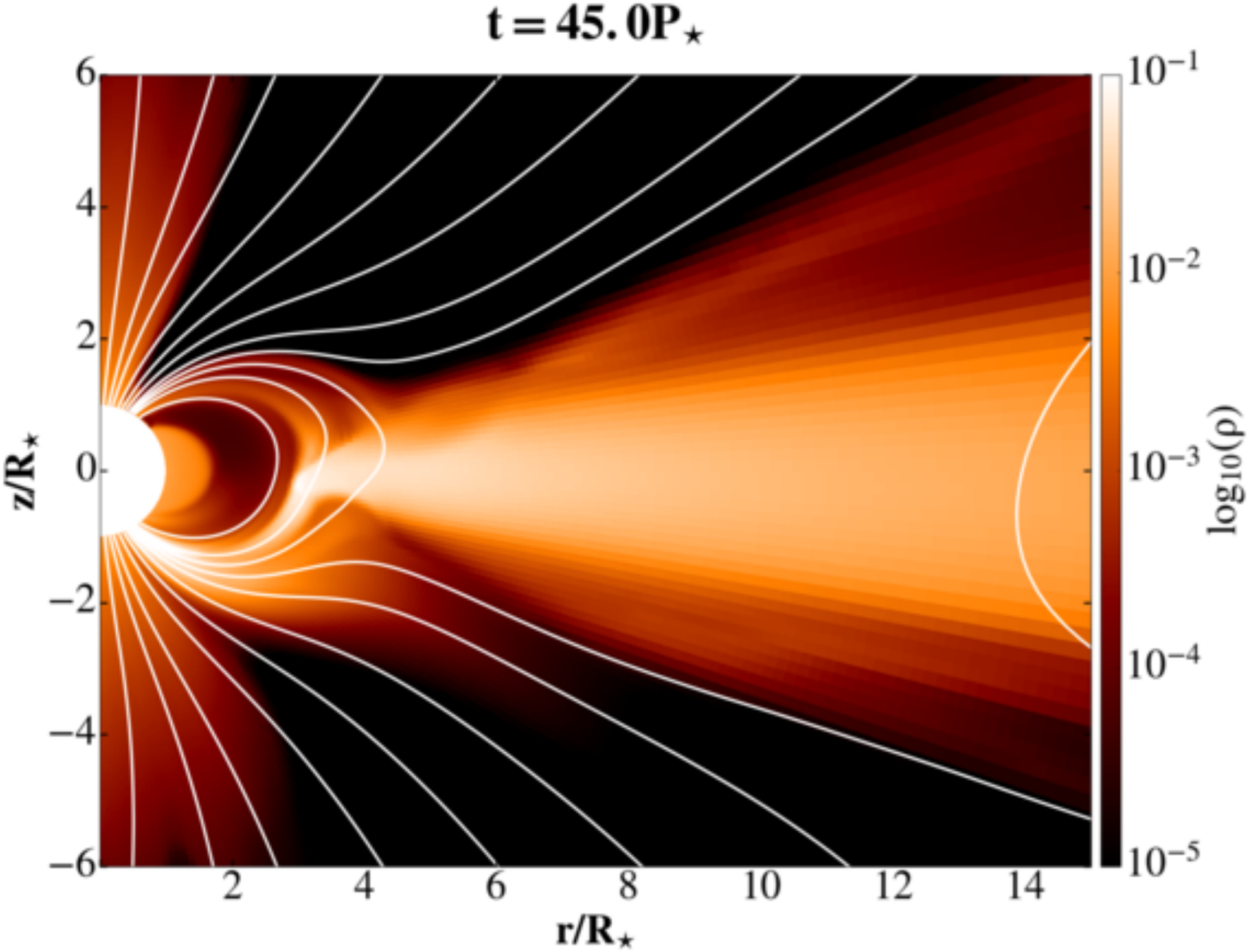}
\includegraphics[width=0.67\columnwidth]{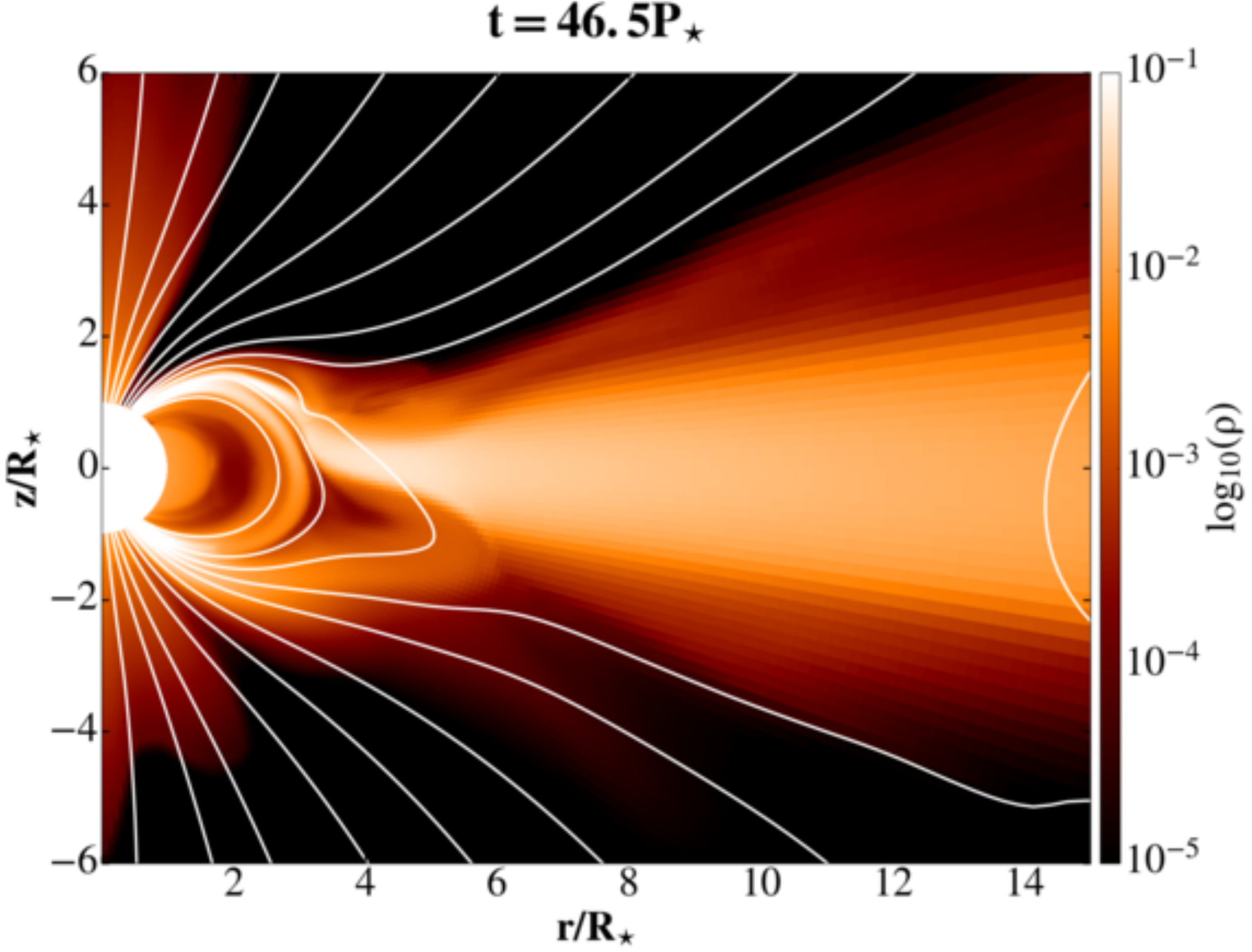}
\includegraphics[width=0.67\columnwidth]{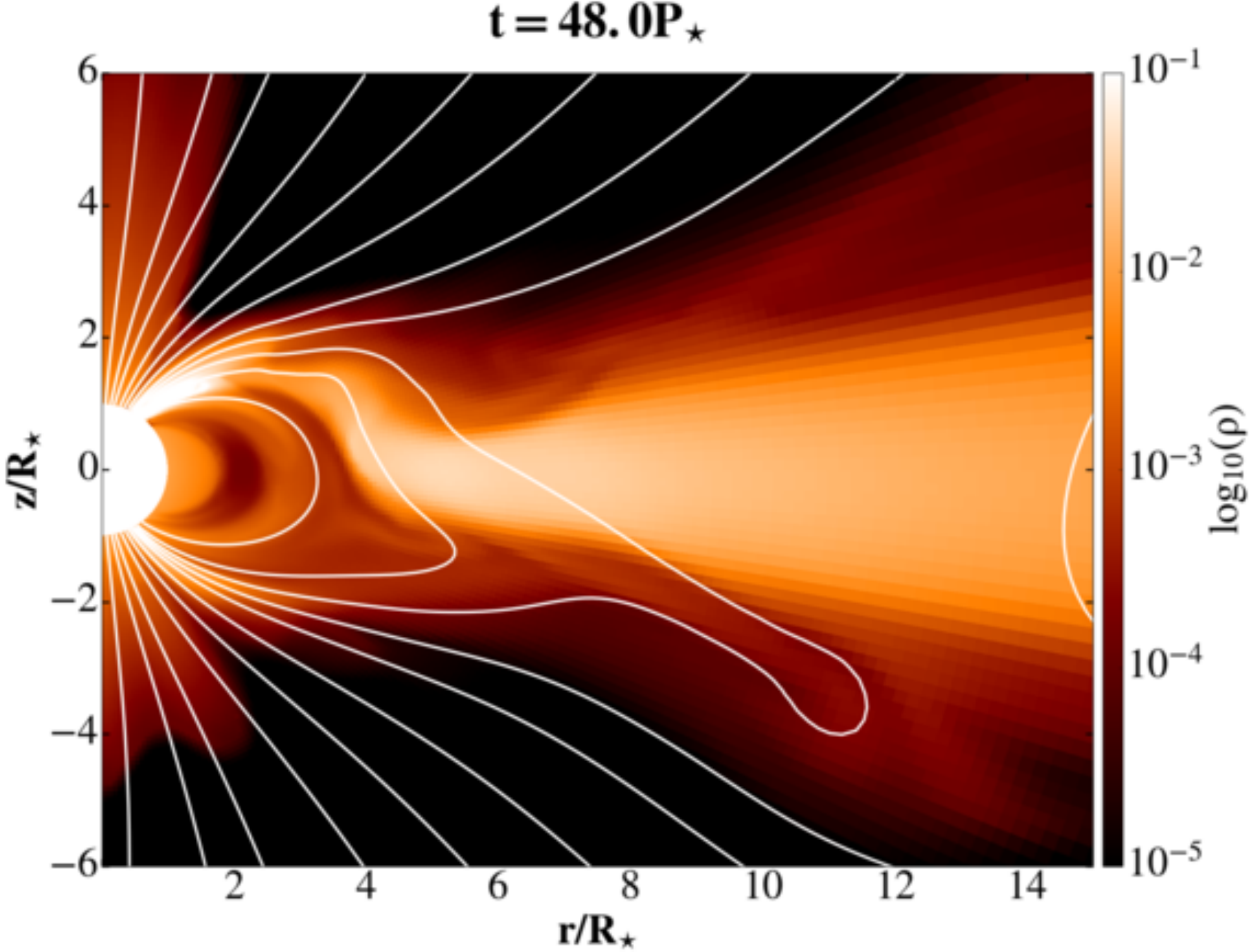}
\caption{A sequence of snapshots from results in the case with
$\theta=[0,\pi]$, in the interval when the switching of the
accretion column from the southern to the northern hemisphere occurs, for
which the intensities are shown in Fig.~\ref{intens0fp}. Colors and lines
have the same meaning as in Fig. \ref{fulpi}.}
\label{fulpi2}
\end{figure*}
Matter inflow onto the surface of a sufficiently magnetized star creates hot
spots, as described in e.g. \cite{ghlamb79}, \cite{cam90}, \cite{shu94}.
Such hot spots are expected in various objects like classical T Tauri stars
\citep{herbst86,johibas95,alenibat02}, cataclysmic variables \citep{wickr91,warn00},
millisecond pulsars \citep{chak03} and X-ray pulsars \citep{bild97}.

Hot spots were obtained in the pioneering series of publications on axisymmetric
two-dimensional (2D) and full three-dimensional (3D) numerical simulations of
star-disk interaction by M. Romanova and collaborators
\citep{R02,Long05,R09,R13}, and in \cite{zf09,zf13}. I confirmed the results
of their 2D numerical simulations in \cite{cpk17,cem18}.

The full 3D simulations as in \cite{R13} would still be impractical for a wide
parameter study needed for a fitting of the observed light curves. Such
simulations are challenging to set-up, and also demand large computational
resources and time. To bridge a gap between the theoretical demands on a code,
and increasingly detailed observations, I extended the axisymmetric 2D
simulations from \cite{cem18} to a complete meridional half-plane. Using this
setup, a 3D model of a star is constructed, with two identical accretion
columns and hot spots positioned diametrically opposite on the stellar surface.
I compute the integrated radiation flux from the surface of such a star, as
observed from various angles.

The obtained intensity curve is closely related to the physical parameters in
the underlying 2D axisymmetric simulation. The free parameters in the simulations
are the stellar rotation rate, magnetic field strength, and the magnetic Prandtl
number. Assuming the plausible dimensions of the stellar hot spots, one can
find the physical parameters of the system best matching the observations. Other
features, like occultation by the disk, or reprocessing of the emission in the
accretion column, or use of the specific filters in the observation, could be
superimposed to the presented solution.

In \S 2 I give a brief description of the numerical simulations set-up, followed
by the results of simulations. The intensity curves in 2D are given in \S 3, and
a 3D model constructed from them is discussed in \S 4. I present a case with
the equatorially symmetric disk in \S 5 and summarize the results in \S 6.

\section{Numerical setup and results of simulations}
I performed simulations of a star-disk system magnetospheric interaction (SDMI) in
a 2D-axisymmetry setup in spherical coordinates, with a logarithmically stretched
grid in the radial direction, spanning 30 stellar radii, and a uniform grid in the
latitudinal direction.

Set-up in the simulations which are used in the model are presented in detail in
\cite{cem18}, which closely follows \cite{zf09}. Here is given only a brief
overview of the solved equations and numerical setup. Viscous and resistive
magneto-hydrodynamic equations are solved with a publicly available code
{\sc pluto} (v.4.1), described in \cite{m07,m12}. The solved equations are,
in the cgs system of units:

\begin{eqnarray}
\frac{\partial\rho}{\partial t}+\nabla\cdot(\rho{\mathbf{v}})=0\\
\frac{\partial\rho{\mathbf{v}}}{\partial
t}+\nabla\cdot\left[\rho{\mathbf{v}}{\mathbf{v}}+\left(P+\frac{{\mathbf{B}}{\mathbf{B}}}{8\pi}\right){\mathbf{I}}-
\frac{{\mathbf{B}}{\mathbf{B}}}{4\pi}-{\mathbf{\tau}}\right]=\rho{\mathbf{g}}\\
\frac{\partial E}{\partial t}+\nabla\cdot\left[\left(E+P+\frac{{\mathbf{B}}
{\mathbf{B}}}{8\pi}\right){\mathbf{v}}-\frac{({\mathbf{v}}
\cdot{\mathbf{B}}){\mathbf{B}}}{4\pi}\right]=\rho{\mathbf{g}}\cdot{\mathbf{v}}\\
\frac{\partial{\mathbf{B}}}{\partial t}+\nabla\times({\mathbf{B}}
\times{\mathbf{v}}+\eta_{\mathrm m}{\mathbf{J}})=0 , 
\end{eqnarray}
where the symbols have have their usual meaning: $\rho$, ${\mathbf{v}}$
and P stand for the matter density, velocity and pressure. The resistivity and
the viscous stress tensor are represented with $\eta_{\rm m}$ and
${\mathbf{\tau}}$, respectively. The gravity acceleration is
${\mathbf{g}}=-\nabla\Phi_{\mathbf{g}}$, with the gravitational potential of
a star with mass $M_\star$ equal to $\Phi_{\mathbf{g}}=-GM_\star/R$. The electric current
is given by the Ampere's law
${\mathbf{J}}=\nabla\times{\mathbf{B}}/4\pi$ and E is the total energy density
$E=P/(\gamma-1)+\rho({\mathbf{v}}\cdot{\mathbf{v}})/2+{\mathbf{B}}\cdot{\mathbf{B}}/8\pi$.
The plasma polytropic index is $\gamma=5/3$.

The disk is set with the initial conditions from \cite{KK00}, with an addition of
the initially non-rotating hydrostatic corona and poloidal magnetic field, shown in
Fig.~\ref{fulpi}. The first of two presented cases in the simulations is with the
computational domain spanning a complete meridional half-plane $\theta=[0,\pi]$, with the
resolution $R\times\theta=(217\times 200)$ grid cells. This simulation extends the
previously obtained star-disk simulation results presented in \cite{cem18}, removing
the disk equatorial boundary condition-the disk mid-plane is now computed
self-consistently in the computational domain. The solution with the disk
equatorial plane given as a boundary condition is presented here as the second
case, where the computational domain spans only a $\theta=[0,\pi/2]$ quadrant of the
meridional half-plane, with the resolution $R\times \theta=(217\times 100)$ grid cells.

The viscosity and resistivity are parameterized by the \citet{ss73} prescription as
$\alpha_{\rm v} c^2/\Omega_{\rm K}$, where $\alpha_{\rm v}$ is a free parameter of
viscosity, smaller than unity, and $\Omega_{\mathrm K}=\sqrt{GM_\star/r^3}$
is the Keplerian angular velocity at the cylindrical radius $r$. A split-field method
is used, in which only changes from the initial stellar magnetic field are evolved in
time \citep{tan94,pow99}. In the simulations is assumed that all the disk heating due
to viscous and Ohmic dissipation is radiated away from the disk-this is why
the heating and cooling terms are removed from the energy equation.
Simulations are still solved in the resistive and viscous regime, because of
the viscosity in the equation of motion and the resistivity in the induction
equation.

After the relaxation from initial conditions, typically after a few tens of the
underlying star rotations, the middle and outer part of the disk in simulations
reach a quasi-stationary state, even if the accretion column still can change
the point of connection to the stellar surface. The results are illustrated in
Fig.~\ref{fulpi2}: after t=45 stellar rotations the column is positioned at the
southern hemisphere and its hot spot produces most of the emitted intensity. After
one and a half rotation, at t=46.5 the column switches from the southern hemisphere
to the northern. There, the new hot spot forms and its intensity is taking over as
the largest contributor in the total intensity. Simultaneously, the southern hot
spot intensity decreases.

Based on the 2D results, I create a 3D model to compute the intensity of radiation
from hot spots, as measured by an observer at infinity, positioned at various angles
with respect to the axis of rotation.

\section{Intensity of radiation along the stellar rim}
\begin{figure*}
\includegraphics[width=0.5\columnwidth]{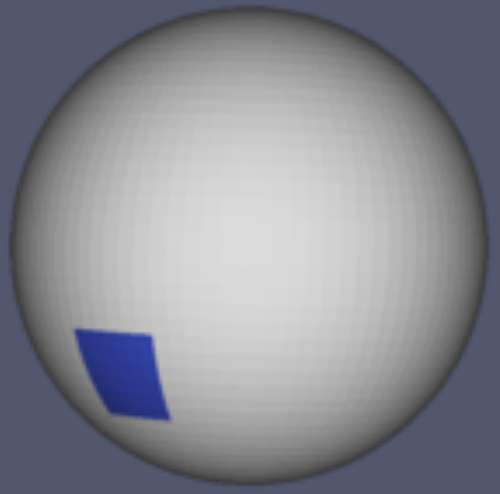}
\includegraphics[width=\columnwidth,height=0.49\columnwidth]{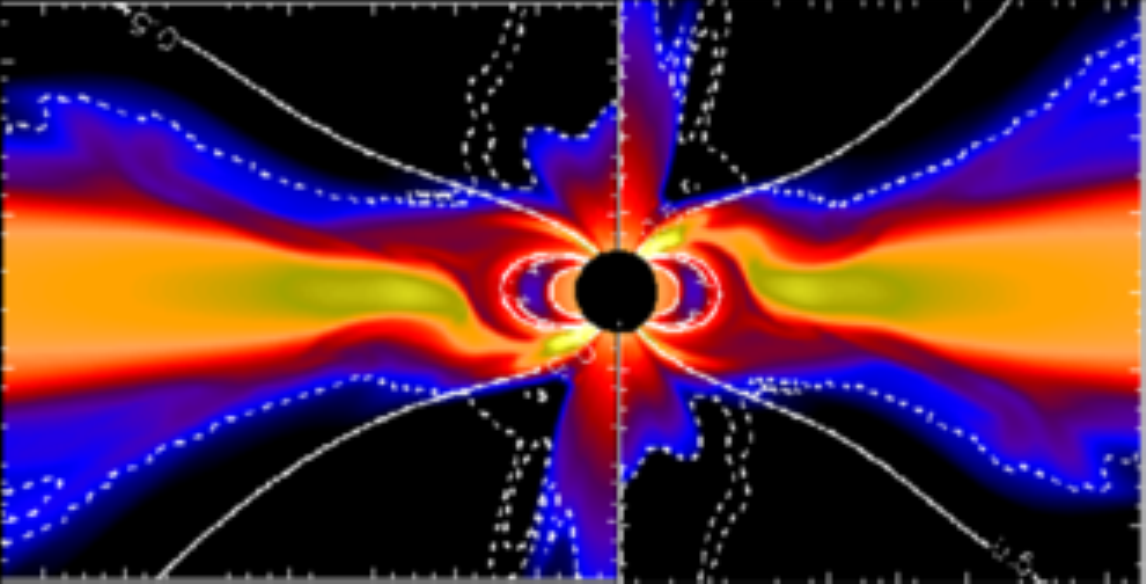}
\includegraphics[width=0.5\columnwidth]{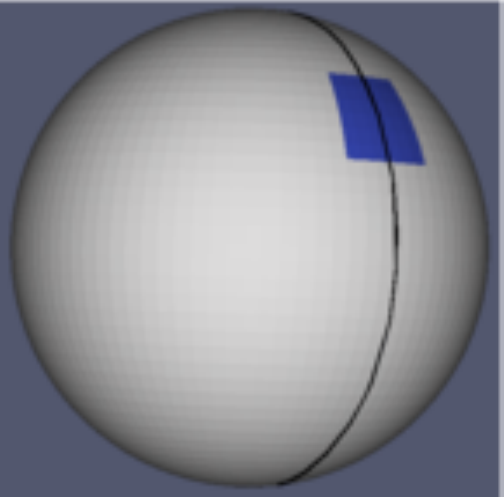}
\caption{A composite of snapshots from the simulations in 2D, shown in the
middle panel, is arranged to show the geometry of the 3D model. The
color and lines meaning is the same as in Fig.~\ref{fulpi}.
In the side panels are shown the corresponding hot spots atop the model sphere,
rotated $\varphi=270^{\circ}$, to show the diametrically opposite hot spots
atop the star.}
\label{model}
\end{figure*}
\begin{figure}
\includegraphics[width=\columnwidth]{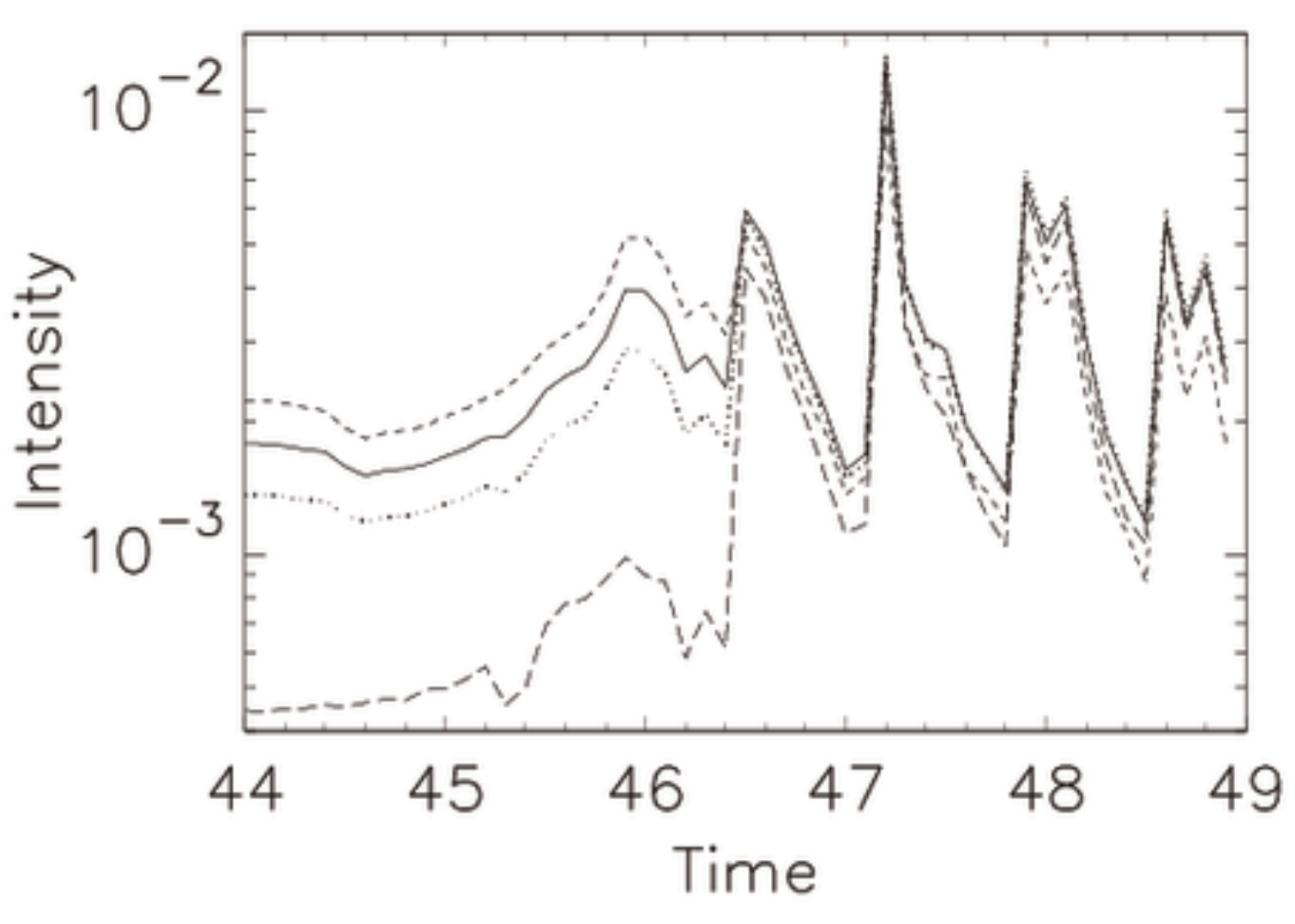}
\includegraphics[width=\columnwidth]{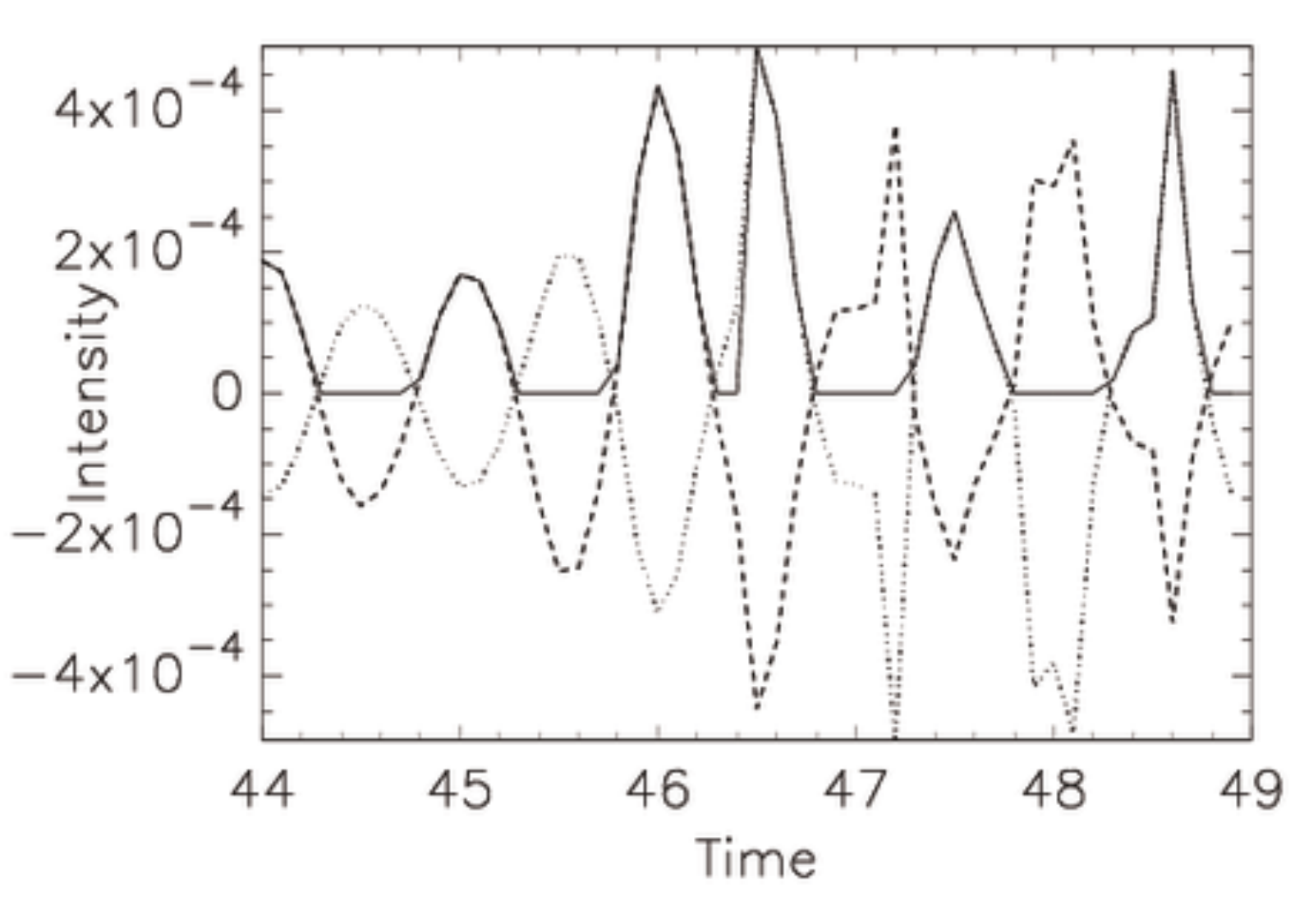}
\caption{Intensity of emission in code units in the simulations in the
$\theta=[0,\pi]$ case. Time is measured in the number of stellar rotations.
Top panel: Intensity for an observer at infinity, integrated along the
stellar rim, which is only one grid cell thick in the azimuthal direction.
The solid, dotted, long-dashed and short-dashed lines represent the
intensities for an observer positioned at a co-latitudinal angle
$\alpha=$15, 30, 60 and 165 degrees, respectively. Bottom panel: Intensity
in a 3D model computed from the result of the 2D simulations shown at the
above panel. The dotted and dashed lines show the intensity of the modeled
hot spots as seen under the angles of $15^\circ$ and $165^\circ$, respectively.
The negative intensity means that the spot is not visible from a given position.
The solid line shows the total intensity measured by an observer at the angle
of $165^\circ$. A shift in phase of the observed light occurs in the time interval
between 45 and 48 rotations of the star. }
\label{intens0fp}
\end{figure}
\begin{figure}
\includegraphics[width=\columnwidth]{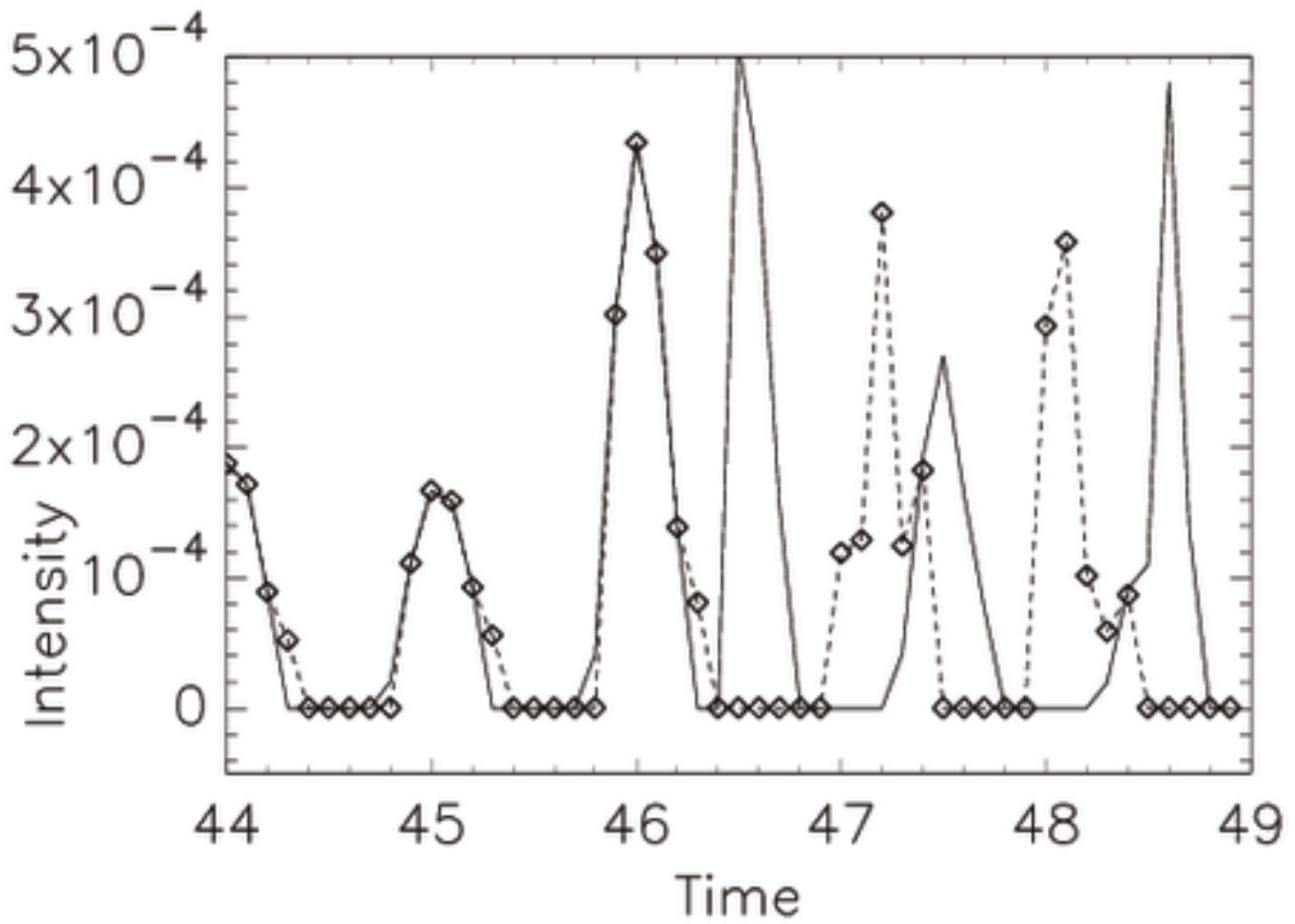}
\caption{Intensity in the simulations in a $\theta=[0,\pi]$ case
when the opposite hot spot is positioned at an angle of 180$^\circ$
(solid line) in the antipodal case, and at 202.5$^\circ$ (dashed line
with diamond symbols) in the non-antipodal case. The intensity is
computed for an observer positioned at an angle of $165^\circ$, measured
from the north pole. The maximum of the intensity is shifted less in
phase in the non-antipodal case.}
\label{intensnoant}
\end{figure}
\begin{figure}
\includegraphics[width=\columnwidth]{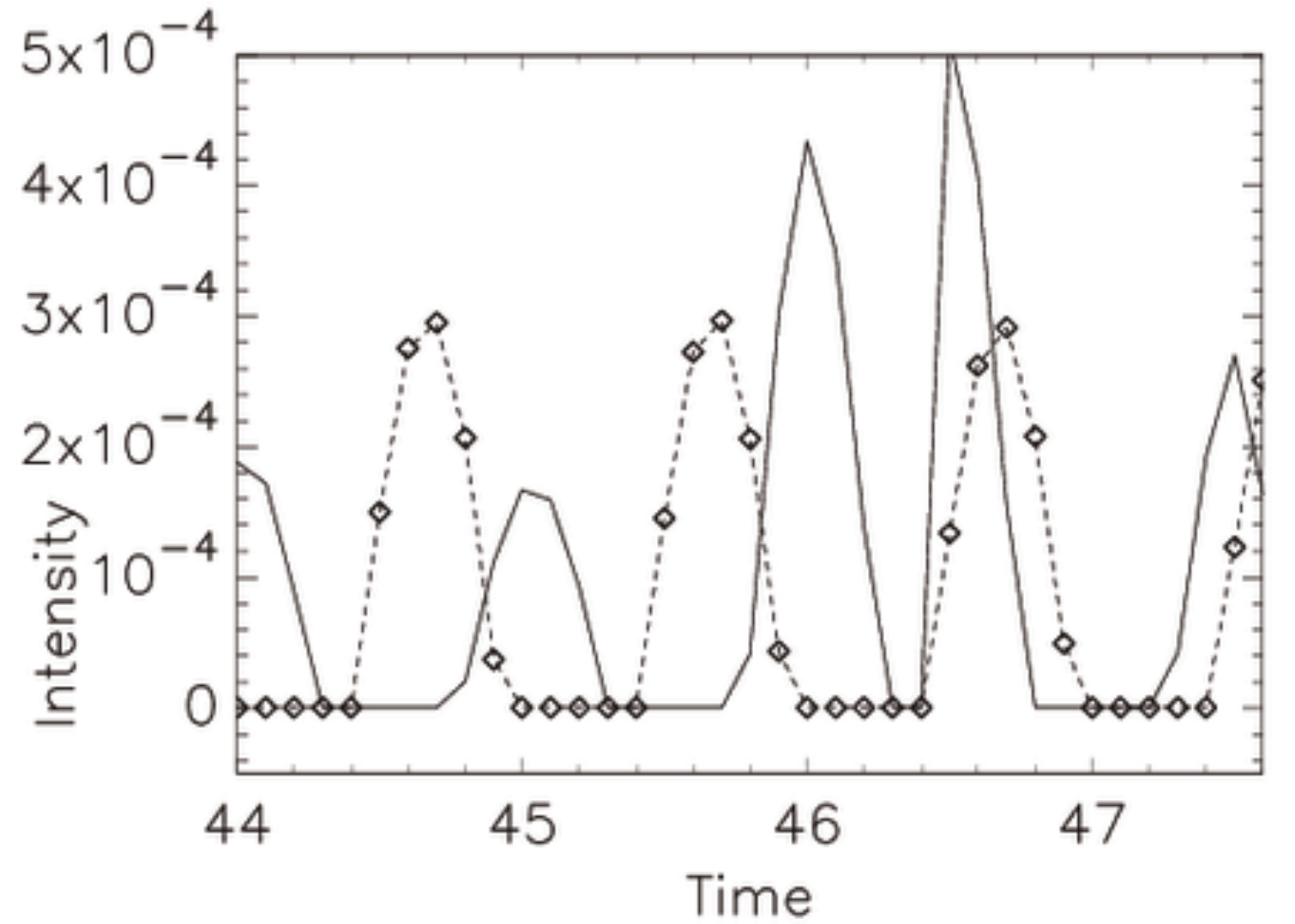}
\caption{Illustration of the difference in the intensity curves in
two simulations in a $\theta=[0,\pi]$ case with the different
physical parameters. The opposite hot spot in the 3D model is
positioned at an angle of 180$^\circ$ in both cases, and the
intensity is computed for an observer positioned at an angle
$165^\circ$, measured from the north pole. The intensity
from Fig.~\ref{intens0fp} in the case with
$\Omega_\star$=$0.1\Omega_{\rm br}$, $\alpha_{\rm m}$=0.7 is
repeated (solid line), along with the intensity in the case with the
$\Omega_\star$=$0.15\Omega_{\rm br}$, $\alpha_{\rm m}$=1 (dashed
line with diamond symbols). The profiles differ in both magnitude
and phase.}
\label{intdifpar}
\end{figure}

\begin{figure}
\includegraphics[width=\columnwidth]{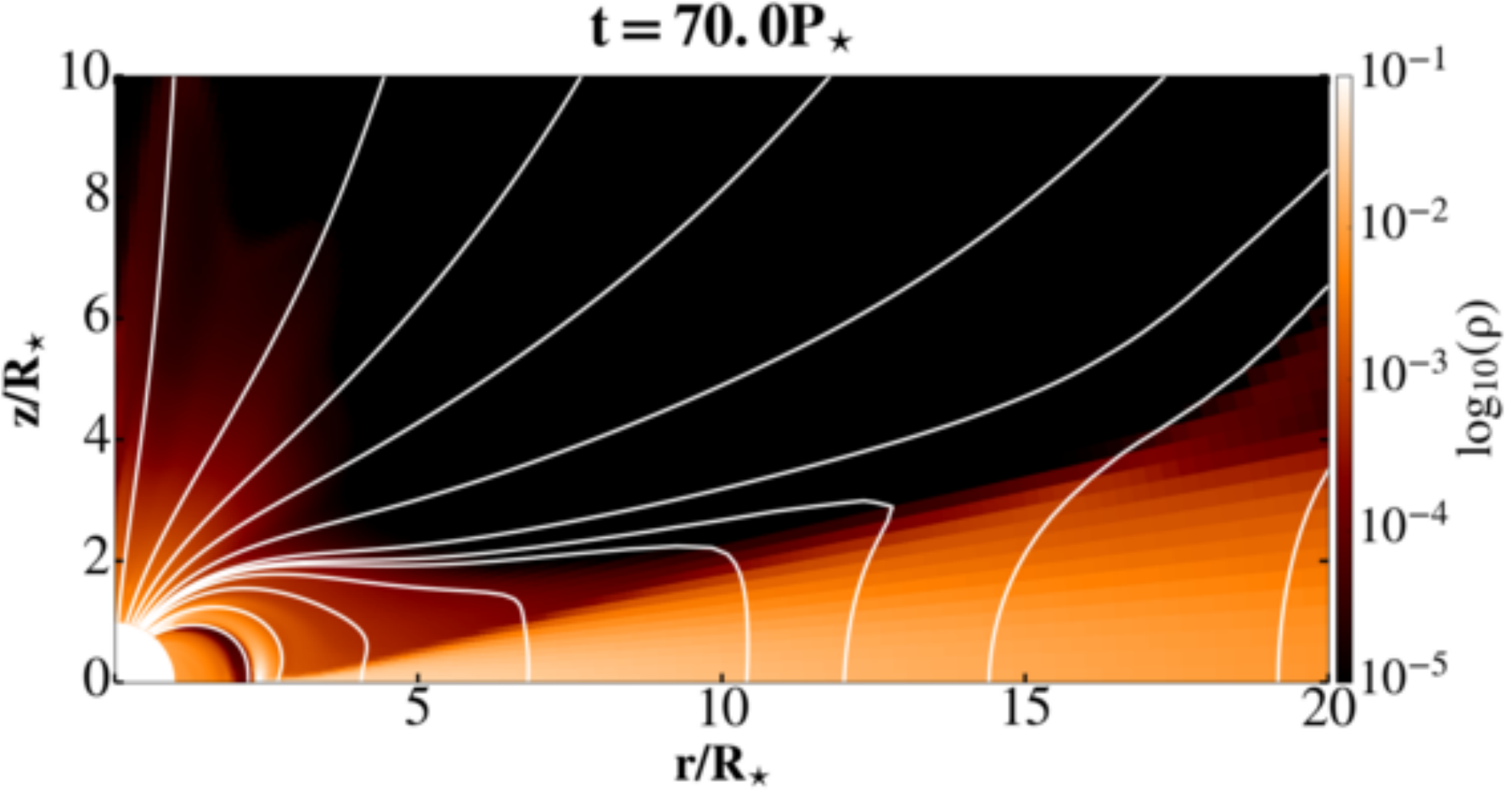}
\caption{Zoom into a snapshot in the simulation in the $\theta=[0,\pi/2]$
case, after T=70 rotations of the central star, when the solution reached
a quasi-stationary state. Lines and colors have the same meaning as in
Fig.~\ref{fulpi}. The initial setup is identical to the first quadrant
in the $\theta=[0,\pi]$ case, shown in Fig.~\ref{fulpi}.
}
\label{halfpi}
\end{figure}
\begin{figure}
\includegraphics[width=\columnwidth]{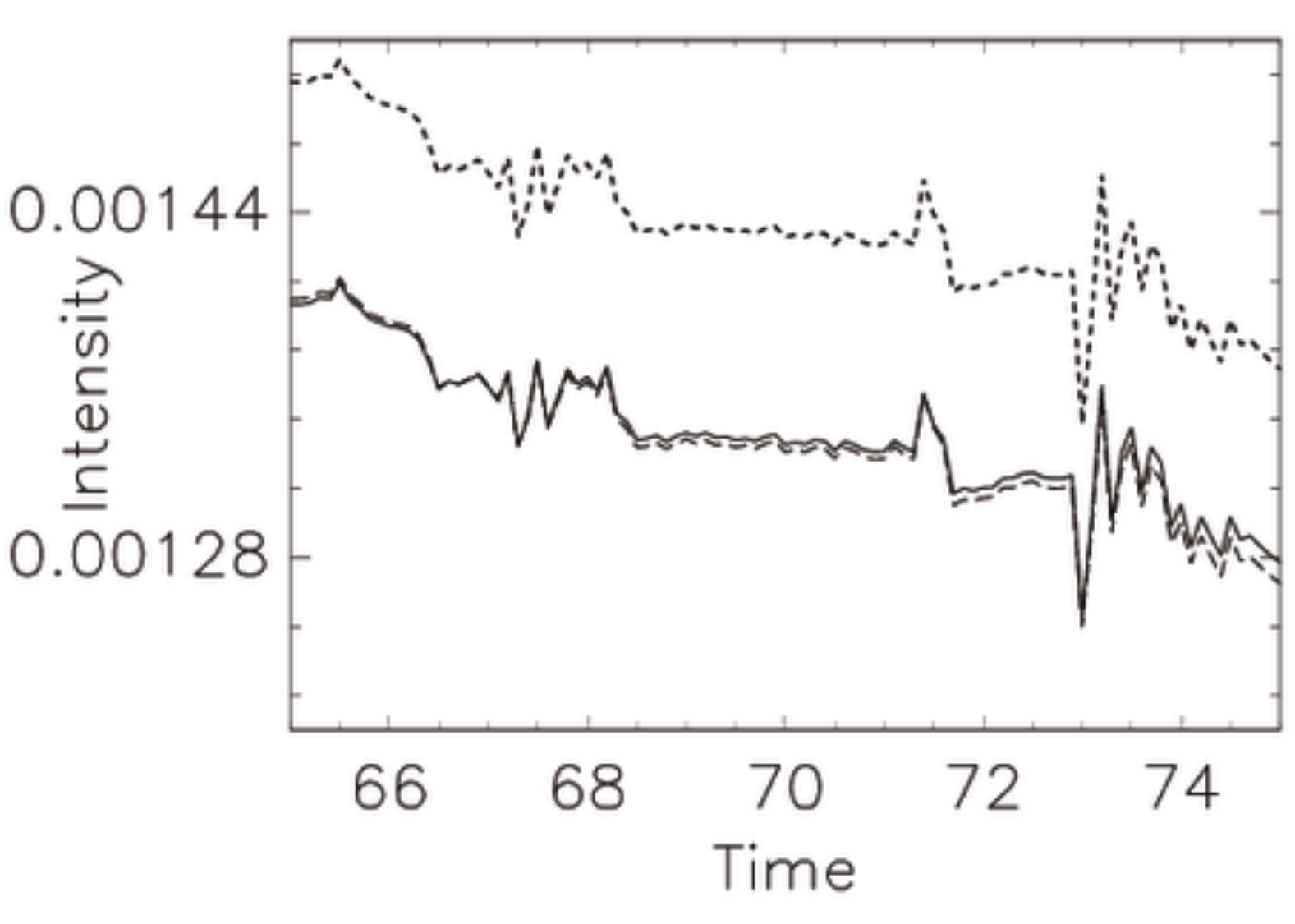}
\includegraphics[width=\columnwidth]{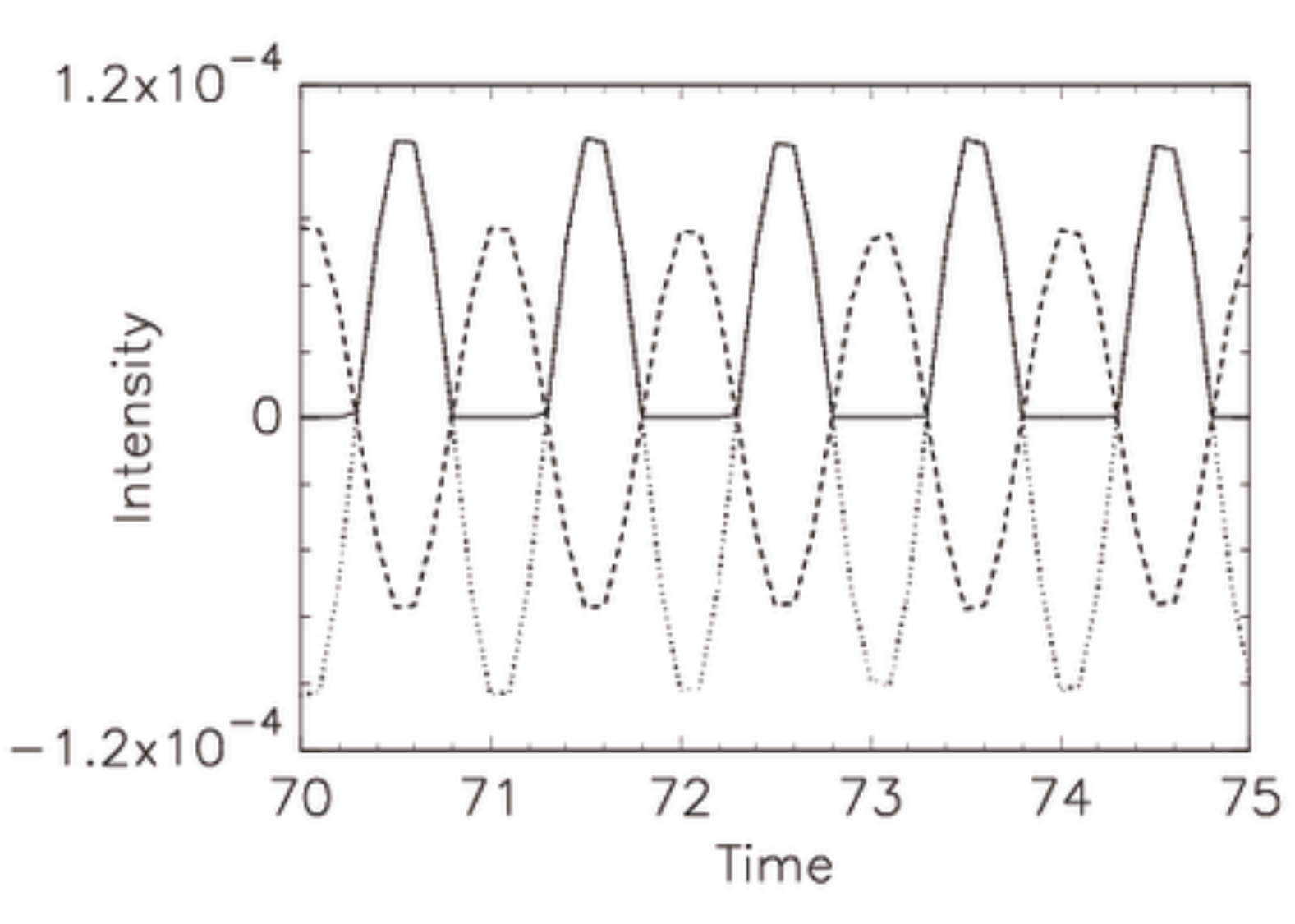}
\caption{Intensity of emission in code units in the simulations in a
$\theta=[0,\pi/2]$ case. Top Panel: Intensity for an observer at infinity,
integrated along the stellar rim. The solid, short-dashed and long-dashed
lines represent the intensities for an observer positioned at a co-latitudinal
angle of $\alpha=$15, 30 and 60 degrees, respectively. Bottom panel: Intensity
in a 3D model computed from the result shown in the above panel. The dotted
and dashed lines show the intensity for the modeled hot spots as seen under
the angles of $15^\circ$ and $165^\circ$, respectively. The solid line shows the
total intensity measured by an observer under the angle of $165^\circ$.
The negative intensity means that the spot is not observable from a given
position.}
\label{intens0hp}
\end{figure}

In a 2D-axisymmetric simulation obtained is only the intensity of radiation
along the line shown on the stellar surface, shown in the right panel in
Fig.~\ref{model}. With this information, I compute the intensity of
emission from a modeled hot spot of the accretion column in the northern
hemisphere. By copying the result into the opposite part of a meridional
plane, I add the antipodal hot spot of the accretion column in the southern
hemisphere, as shown in the left panel.

The intensity of radiation emitted along the stellar rim is
\begin{eqnarray}
I_R=\frac{1}{\pi}\int_{\cos(\alpha-\theta)>0}F({\mathbf{R}})\cos\alpha~dS\ ,
 \mathrm{where}\nonumber \\ 
F({\mathbf{R}})=\rho {\rm v}_r(\frac{1}{2}{\rm v}^2+\frac{\gamma}{\gamma-1}\frac{P}{\rho}) \nonumber
\end{eqnarray}
is the intensity of the radiation in the direction of the observer at infinity,
positioned at the co-latitudinal angle $\alpha$ with respect to an element of the
surface $dS$ - see e.g. \cite{R04}. In this intensity, the stellar radiation or
any obscuring effects from the disk or column is not accounted for, and the
emission from the column above the stellar surface is not included.

To better illustrate the light curve modeling procedure, I choose a particular
example of an uneven distribution of the intensity peaks in a 2D simulation.
Such a distribution is caused by the switching of the accretion column
from one stellar hemisphere to the other, shown in Fig.~\ref{fulpi2}. This
could help to model some of the unexplained dips, or "occultations" in the
observed light curves, as in e.g. \cite{siw14}.

In Fig.~\ref{intens0fp}, shown is a change in time in the intensity curve
integrated along the stellar rim, obtained in the 2D simulation with a star
rotating with 1/10 of the stellar breakup rate, viscosity and
resistivity parameters equal to 1 and 0.7, respectively, and the stellar
dipole magnetic field of 500~G. Time is measured in the number of stellar
rotations. The observed light curve indicates that the column switched
hemisphere. Before the switch, until t=46.5 the accretion column is steadily
positioned in the southern hemisphere and the largest part of the emission is
radiated into the portion of space with $\alpha>90^\circ$ (shown in the
short-dashed line). After the switch of the column to the northern hemisphere,
the weakening southern hot spot still contributes to the total intensity, but
more of the intensity is emitted from the newly formed northern hemisphere hot
spot into the $\alpha<90^\circ$ part of the computational domain (shown in the
solid, dotted and long-dashed lines).

\section{Intensity of radiation in 3D model}
A hot spot in simulations is of a $\Delta\theta$ span in the latitudinal
direction. To obtain a 3D model, an azimuthal extension of the hot
spot is added, with the assumption that it is equal to the latitudinal one
so that $\Delta\varphi=\Delta\theta$. The obtained hot spot is shaped as a
part of the spherical surface shown in Fig.~\ref{model}. To complete
the 3D model with two antipodal accretion columns, such a hot spot is copied
into a diametrically opposite hot spot on a sphere.

By rotating a model star with hot spots, the radiation intensities can be obtained
for observers positioned at different angles with respect to the axis of rotation.
The intensity radiated from the surface of a rotating star towards an observer
positioned at the angle $\alpha$, measured from the north pole, and positioned
in the plane with the zero azimuthal angle $\varphi=0$ is
\begin{equation}
 I=\int_{\varphi}\cos\varphi d\varphi\int_0^\pi R^2 F({\mathbf{R}})\sin\theta\cos
(\alpha-\theta)d\theta \ . \nonumber
\end{equation}
The results of computation are shown in the bottom panel in Fig.~\ref{intens0fp}.
Switching of the accretion column between the stellar hemispheres is present in the
modeled intensity curve as a shift of phase in the measured peaks. It follows the
visibility of the hot spots because of the stellar rotation.

When analyzing the data from the observations, with a suitable set of parameters
in the described model one could modify the constraints to match the observed light
curve. The shape and position of the spots could be shifted in both latitude and
azimuth. The azimuthal extension of each hot spot could be chosen individually.

An example of modification is a non-antipodal positioning of the
opposite hot spot. In Fig.~\ref{intensnoant}, with a dashed line and indicated
also with the diamond symbols, is shown the intensity in a case when the
opposite hot spot is positioned at an angle of 202.5$^\circ$, instead of
180$^\circ$. The shape of this light curve is also modified. For comparison,
the intensity in the antipodal case is also shown. The maximum of the intensity
is less shifted in phase after the column switches the hemisphere than in
the antipodal case.

The obtained 3D model is dependent on physical parameters in the underlying
2D numerical simulation. In Fig.~\ref{intdifpar} is shown an illustration,
where a shift in phase in light curve is shown between the two simulations
with the different stellar rotation rates. The light curves are computed
in the cases with the different resistivities in the disk, so that the obtained
curves are also of the different shape and magnitude-with the solid line is
repeated the intensity in the case from Fig.~\ref{intens0fp}, and with the dashed
line in the case with the stellar rotation rate 1/15 of the breakup rate, and
the resistivity parameter $\alpha_{\rm m}$=1. The stellar magnetic field
and viscosity in two simulations are the same, 500~G and $\alpha_{\rm v}$=1,
respectively. In the simulation with a larger disk resistivity, there is no
switch in the position of the accretion column, it remains at the northern stellar
hemisphere throughout the simulation.

\section{A case with equatorially symmetric disk}
Simulations are often performed in only the $\theta=[0,\pi/2]$ part of
the meridional plane, with an assumption of the equatorial symmetry in the
disk. To create a 3D model in such cases, I repeated the simulation with the
same parameters in a new domain, spanning only over the northern quadrant in the
meridional plane. A snapshot in a quasi-stationary solution is
shown in Fig.~\ref{halfpi}. The obtained intensity along the rim in a 2D simulation
in this case is shown in the top panel in Fig.~\ref{intens0hp}. It is of a similar
order of magnitude as in the $\theta=[0,\pi]$ case. Since now only the intensity
along a northern hemisphere part of the meridian line is obtained in the
simulation, the hot spot can not switch the hemisphere. In the model, I assumed
that there is no contribution to the intensity along the other, not simulated
half of the latitudinal arc.

Creating a 3D model with the antipodal hot spot from this result, I obtain the
intensity shown in the bottom panel in Fig.~\ref{intens0hp}. Since there is no
contribution from the switching of the position of the hot spot from one to the
other hemisphere, there is no change in the position of the dip of the intensity
curve. An observer would see the contribution to the intensity only from one hot
spot at a time. 

Such an intensity curve is also observed in the case when only one accretion
column forms from the accretion disk onto the stellar surface. The
intensity measured by an observer under the angle of $165^\circ$ is shown
with the solid line in Fig.~\ref{intens0hp}.

\section{Summary}
I compute the intensity of emission from a hot spot created by the accretion
column infalling onto a stellar surface in the numerical simulations of
star-disk magnetospheric interaction. From the results of 2D simulations,
with minimal additional assumptions is computed a full 3D model.

In a descriptive example, I show how a shift in phase of the intensity peaks can
be explained by switching of the accretion column from one stellar hemisphere to
the other. The obtained intensity is closely related to the geometrical and
physical parameters of the underlying star-disk numerical simulation.

One case of a difference in results with modification of the geometry in 3D
model is illustrated in an example, in which the opposite hot spot on the
stellar surface is not placed exactly in the antipodal position. The shape
and phase of the intensity maximum change.

To further illustrate the versatility of the model, I compare the modeled
intensity curves with a modification in the physical parameters in the
underlying numerical simulation. Both the shape and phase of the
resulting intensity curves in the full 3D model are modified with such a
change. I also show the results in the cases when a hot spot does not
switch the stellar hemisphere, or when there is only one accretion column
hot spot at the stellar surface.

In the presented models, assumed are hot spots of the same shape and area.
With the appropriate modifications, intensity from hot spots of
different extensions can be computed. With an addition of the observational
filter responses and occultation patterns with the disk and column, the model
could account for the different geometries of the system and the effects
of a medium through which the radiation passes on the way to an observer. 

\section*{Acknowledgements}
M\v{C} developed the setup for star-disk magnetospheric simulations while in
CEA, Saclay, under the ANR Toupies grant with A.S. Brun. Work in NCAC Warsaw
is funded by a Polish NCN grant no. 2013/08/A/ST9/00795 and a collaboration with
the Croatian STARDUST project through HRZZ grant IP-2014-09-8656 is acknowledged.
I thank IDRIS (Turing cluster) in Orsay, France, ASIAA/TIARA (PL and XL
clusters) in Taipei, Taiwan and NCAC (PSK cluster) in Warsaw, Poland, for
access to Linux computer clusters used for high-performance computations. The
{\sc pluto} team is thanked for the possibility to use the code. V. Parthasarathy,
A.D. Bollimpalli and NCAC summer students C. Turski and F. Bartoli\'{c} are
thanked for developing the Python scripts for visualization, and M. Siwak for
the initial motivation for this work.

\bsp	
\label{lastpage}

\begin{thebibliography}{}
\bibitem[Alencar \& Batalha(2002)]{alenibat02}
Alencar, S.H.P., \& Batalha, C., 2002, \apj, 571, 378
\bibitem[Bildsten(1997)]{bild97}
Bildsten, L., et al., 1997, \apjs, 113, 367
\bibitem[Camenzind(1990)]{cam90}
Camenzind, M., 1990, Rev. Mod. Astron., 3, 234
\bibitem[\v{C}emelji\'{c} et al.(2017)]{cpk17}
\v{C}emelji\'{c}, M., Parthasarathy, V. \& Klu\'{z}niak, W., 2017,
JPhCS, 932, 012028
\bibitem[\v{C}emelji\'{c}(2018)]{cem18}
\v{C}emelji\'{c}, M., 2018, arXiv:1811.02808, submitted to A\&A
\bibitem[Chakrabarty et al.(2003)]{chak03}
Chakrabarty, D., Morgan, E.H., Muno, M.P., Galloway, D.K., Wijnands, R., Van
Der Klis, M., \& Markwardt, G.B., 2003, Nature, 424, 42
\bibitem[Ghosh \& Lamb(1997)]{ghlamb79}
Ghosh, P., \& Lamb, F.K., 1997, \apj, 234, 296
\bibitem[Herbst et al.(1986)]{herbst86}
Herbst, W., et al., 1986, \apj, 310, L71
\bibitem[Johns \& Basri(1995)]{johibas95}
Johns, C.M., \& Basri, G., 1995, \apj, 449, 341
\bibitem[Klu\'{z}niak \& Kita(2000)]{KK00}
Klu\'{z}niak, W., Kita, D., 2000, arXiv:astro-ph/0006266
\bibitem[Mignone et al.(2007)]{m07}
Mignone, A., Bodo, G., Massaglia, S., Matsakos T., Tesileanu O.,
 Zanni C., Ferrari A., 2007, \apjs, 170, 228
\bibitem[Mignone et al.(2012)]{m12}
Mignone, A., Zanni, C., Tzeferacos, P., van Straalen, B., Colella,
P., and Bodo, G., 2012, \apjs, 198, 7
\bibitem[Powell et al.(1999)]{pow99}
Powell, K. G., Roe, P. L., Linde, T. J., Gombosi, T. I., \& De Zeeuw, D. L.
1999, J. Comput. Phys, 154, 284
\bibitem[Romanova et al.(2002)]{R02}
 Romanova, M.M., Ustyugova, G.V., Koldoba, A.V., Lovelace, R.V.E., 2002,
\apj, 578, 420
\bibitem[Romanova et al.(2004)]{R04}
 Romanova, M.M., Ustyugova, G.V., Koldoba, A.V., Lovelace, R.V.E., 2004,
\apj, 610, 920
\bibitem[Long et al.(2005)]{Long05}
Long, M., Romanova, M.M., \& Lovelace, R.V.E., 2005,
\apj, 634, 1214
\bibitem[Romanova et al.(2009)]{R09}
 Romanova, M.M., Ustyugova, G.V., Koldoba, A.V., Lovelace, R.V.E., 2009,
\mnras, 399, 1802
\bibitem[Romanova et al.(2013)]{R13}
 Romanova, M.M., Ustyugova, G.V., Koldoba, A.V., Lovelace, R.V.E., 2013,
\mnras, 430, 699
\bibitem[Shakura \& Sunyaev(1973)]{ss73}
Shakura, N.I., Sunyaev, R.A., 1973, A\&A, 24, 337
\bibitem[Shu et al.(1994)]{shu94}
Shu, F.H., Najita, J., Ostriker, E.C., Wilkin, F., Ruden, S.P., \& Lizano,
S., 1994, \apj, 429, 781
\bibitem[Siwak et al.(2014)]{siw14}
Siwak, M., Rucinski, S.M., Matthews, J.M., Guenther, D.B., Kuschnig, R.,
Moffat, A.F.J., Rowe, J.F., Sasselov, D., Weiss, W.W., 2014,
\mnras, 444, 327
\bibitem[Tanaka(1994)]{tan94}
Tanaka, T. 1994, J. Comput. Phys., 111, 381
\bibitem[Warner(2000)]{warn00}
Warner, B., 2000, PASP, 112, 1523
\bibitem[Wickramasinghe et al.(1991)]{wickr91}
Wickramasinghe, D.T., Wu, K. \& Ferrario, L., 1991, \mnras, 249, 460
\bibitem[Zanni \& Ferreira(2009)]{zf09}
Zanni, C., Ferreira, J., 2009, A\&A, 512, 1117
\bibitem[Zanni \& Ferreira(2013)]{zf13}
Zanni, C., Ferreira, J., 2013, A\&A, 550, A99
\end{thebibliography}
\end{document}